\documentclass[letterpaper,12pt]{article}

\usepackage{setspace}
\usepackage[margin=1in]{geometry}
\usepackage[affil-it]{authblk}
\usepackage{amsmath,amssymb,amsthm}
\usepackage{parskip}
\usepackage{graphicx}
\usepackage[round]{natbib} 

\usepackage{color,hyperref,xcolor}
\hypersetup{colorlinks=true,urlcolor=blue, citecolor=purple}
\usepackage{cleveref}

\newtheoremstyle{general}
{3mm} 
{3mm} 
{\it} 
{} 
{\bfseries} 
{.} 
{.5em} 
{} 

\newcommand{\norm}[1]{\left\| #1 \right\|}
\newcommand{\cov}{\mathrm{Cov}}

\newcommand{\eps}{\varepsilon}

\newcommand{\given}{\hspace{2pt}\vert\hspace{2pt}}

\newcommand{\dash}{^{\prime}}

\newcommand{\ind}{\mathbb{I}}

\newcommand{\ber}{\mathrm{Bernoulli} \ }

\theoremstyle{general}

\numberwithin{equation}{section}
\numberwithin{theorem}{section}

\begin{document}

\title{Spatial modeling of shot conversion in soccer to single out goalscoring ability}

\author{Soudeep Deb
  \thanks{Email: \texttt{sdeb@uchicago.edu}}}
\affil{Department of Statistics,\\ The University of Chicago,\\ 5747 S Ellis Ave,\\ Chicago, IL, USA. 60615 }

\author{Debangan Dey
  \thanks{Email: \texttt{ddey1@jhu.edu}}}
\affil{Department of Biostatistics, \\ Johns Hopkins Bloomberg School of Public Health,\\ 615 N Wolfe Street,\\ Baltimore, MD, USA. 21205}



\maketitle

\newpage

\begin{abstract}
\noindent Goals are results of pin-point shots and it is a pivotal decision in soccer when, how and where to shoot. The main contribution of this study is two-fold. First, we show that there exists high spatial correlation in the data of shots across games. Then, we introduce a spatial process in the error structure to model the probability of conversion from a shot depending on positional and situational covariates. The model is developed using a full Bayesian framework. Next, based on the proposed model, we define two new measures that can appropriately quantify the impact of an individual in soccer, by evaluating the positioning sense and shooting ability. As a practical application, the method is implemented on Major League Soccer data from 2016/17 season. 

\vspace{0.1in}

\end{abstract}

\newpage

\section{Introduction}
\label{sec:introduction}

Association football or Soccer is arguably the most popular sport in the world. Along with the excitement involved with the international fixtures, millions of viewers across the world also watch the club football on a regular basis. And with the ever-growing competitive nature of the sport, the field of soccer analytics is becoming more important with every passing day. The most interesting aspect about soccer is that it is a dynamic game and a team's success relies both on team strategies and individual player contributions. Hence, it is generally more difficult to develop sophisticated statistical methods for soccer as compared to, say baseball, where events are more discrete. 

In soccer, the soccer-ball is certainly like the principal atom in the game, around which the players on the pitch trace their path. Here, a shot gives momentum to the atom of the game, read soccer ball, to reach its intended destination, the back of the net. In this aspect, shooting ability of a player has always been a key factor in building team strategies, as the outcome of a game is decided based on the number of goals scored by the two teams. In this paper, we address this particular problem in detail. Our primary objective is to develop a new and more efficient model that can successfully capture the effect of different positional and situational variables on how well a player can convert a shot to a goal. On a related note, it is  also immensely important to identify players who have great positioning sense and can successfully convert the chances to score more frequently than others. Following the model we develop, we further define a measure which captures the information about a player's positioning sense and shooting efficiency appropriately. 

A simple way of looking at this problem is to calculate the proportion of shots one converts successfully, which works around the assumption that probability of scoring a goal is the same for all shots taken by a certain player. While it gives some idea about the efficiency of a striker, it has its shortcomings and would fail to serve the purpose from many aspects. This crude measure only gives the percentage of shots converted successfully, but it does not tell anything about the players' positioning senses. Besides, this measure fails to identify difficulty of chances. For example, a player scoring 2 goal from 4 shots, all from several yards out of the box is considered equally efficient as another who scores 1 from 2 shots from inside the box, and that is definitely not the case. Naturally, one should take into account many other factors - location, time, type of play, how the shot was taken, opponents etc - to investigate the conversion rates of soccer players.

While the problem of predicting goals has not been overwhelmingly explored in soccer, there are some studies that have tried to predict the outcome of an average shot of an average player from a particular position on the pitch in a particular situation. Of course, there are limitations as we do not know the shot speed, the position of defenders and the exact goalkeeper position but it is worthy to check how much variance we can explain from the existing covariates which, at the first look, seem to be really important. Current researchers in this field have taken various routes to assign a goal-scoring probability to a particular shot and finally, worked on to develop an Expected Goals (ExpG) metric for a team or a player.

In general, there are different types of discrete regression models that people have used so far. \citet{goddard05} used bivariate Poisson and ordered probit regression models to forecast number of goals scored and conceded in soccer matches based on different covariates. The author here analyzed data from English football and learned that the number of goals in a match depends mostly on the recent home performances of the home side and the recent away performances of the away side. In another match-level study, \citet{williams11} studied the effects of altitude on the result of soccer matches in South America. Taking a regression-based approach for change in altitude of the away side, they showed that the win probability decreases significantly for an away team traveling up, while traveling down does not have a significant impact on the results.

Player-related or shot-related approaches have been studied by a few authors as well. One of the most popular methods in this regard is a study in \citet{asa16}. Analyzing data from the Major League Soccer (MLS), the authors showed that the shot conversion rates significantly depend on many factors, including the distance, angle and the body part used when a shot is taken. Simple logistic regression techniques were the main components of this study. In an earlier paper, \citet{moura07} dealt with data from Brazilian first division championship matches and analyzed shots to goal strategies based on the field position from where the ball was collected and the length of the passing sequence leading to a goal. Contrary to the popular belief, they found that the ball possessions resulting in shots to goal usually start in the defensive field of the scoring teams. So far as the passing sequences are concerned, this study showed that either short (2 to 5) or very long (10 or more) sequences significantly improve the conversion rates. This second finding corroborated what \citet{hughes05} found before. 

Finally, we cite two works that used the situation in a small time-window just preceding the event of interest. First, \citet{jordet13} used English Premier League data to investigate how visual exploratory behaviors before receiving the ball affect the probability of completing the next pass. They found that the players who show extensive visual exploratory behaviors right before receiving a ball are more successful in completing a pass. In another recent study, \citet{lucey14} analyzed the spatiotemporal patterns of the ten-second window of play before a shot was fired. They extracted some strategic features from the data and used them to present a method that estimates the likelihood of a team scoring a goal. This study showed that not only is the game phase (corner, free-kick, open-play, counter attack etc.) important, but the features such as defender proximity, interaction of surrounding players and speed of play coupled with the shot location also play an impact on determining the likelihood of a team scoring a goal. A few more relevant discussions can be found in \citet{leitner10}, \citet{clark13}, \citet{lasek13} and \citet{brooks16}.

In this study, we want to look at the problem of modeling shot conversion rates from a slightly different point of view. While each of the above models has its advantages, there are some limitations too. In order to perform the last two methods, one would need the player tracking data from some reputable source and that requires a considerable amount of money. On the other hand, usual binary regression based approaches often show a lack of fit and fail to have good predictive abilities. This has been discussed more in \Cref{sec:motivation}.

Our primary goal here is to build a model that fits the data well, identifies the factors significantly affecting the shot conversion rate and provides an estimate of the probability of scoring a goal if the location and other details are given. In this regard, we use a spatially correlated error process in the binary regression model. Another key contribution of this paper is to quantify the impact of an individual in soccer. In particular, we define two new measures that evaluate the positioning senses and the goal-scoring abilities of the players. As a practical application, we would implement our model, in conjunction with the measures we devise, to Major League Soccer (MLS) data from 2016/17 season.

The paper is organized as follows. A brief description of the data and some exploratory analysis are provided in the next section. In \Cref{sec:motivation}, we provide motivations behind our idea of including a spatial correlation term in a probit regression model. The proposed method and the measures to evaluate shooters' efficiency are presented in \Cref{sec:methods}. Detailed analysis of the data using the proposed model is carried out in \Cref{sec:results} while the final section includes a discussion, along with some important concluding remarks.

\section{Data}
\label{sec:data}

\begin{figure}[!thb]
\begin{center}
\includegraphics[width=\textwidth,keepaspectratio]{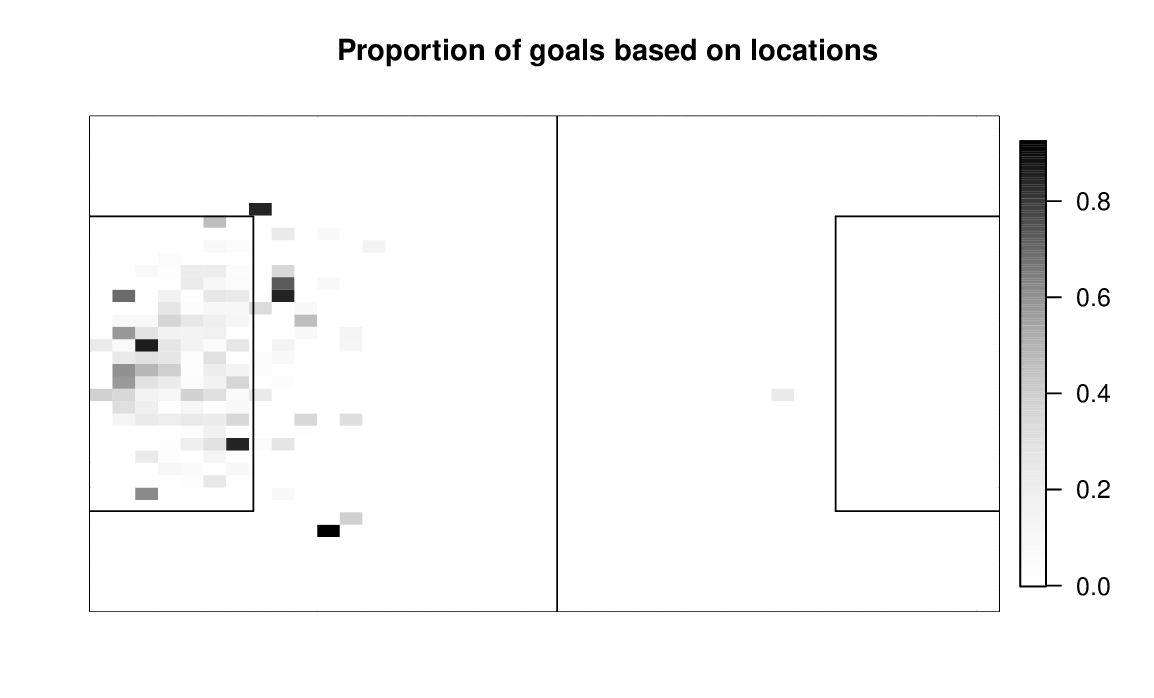}
\end{center}
\caption{Heat map of the full data: Total number of attempts were 3957, among which 482 were goals and 3475 were misses. Each pixel represents the proportion of goals scored from that position, color coded on a scale from 0\% (white) to 100\% (black).}
\label{fig:heatmap}
\end{figure}

Throughout this study, we use the data for the first half of the MLS 2016/17 season. Since the focus is on the shots to goal conversions, relevant information for all shots (excluding penalties, own goals and cases with some missing data) and their outcomes - encoded as {\it goal} or {\it miss} - are recorded. But, we choose to exclude cases when a shot was taken from beyond the half line, for often these shots are simply the results of some unusual mess-up of the opponent side or the result of a moment of some striker's presence of mind. In other words, a player will not take a shot from beyond the center line unless he is handed a golden opportunity and naturally, such a shot has a higher probability of being converted. So, including these instances are likely to affect our conclusions. After removing these events, the data comprised of 3957 attempts at goal, among which 482 were converted successfully (approximately 12.18\% of the total number of shots). A heat map is shown in \Cref{fig:heatmap} to show the proportions of goal scored from different locations in the field.

\begin{figure}[!hbt]
\begin{center}
\includegraphics[width=\textwidth,keepaspectratio]{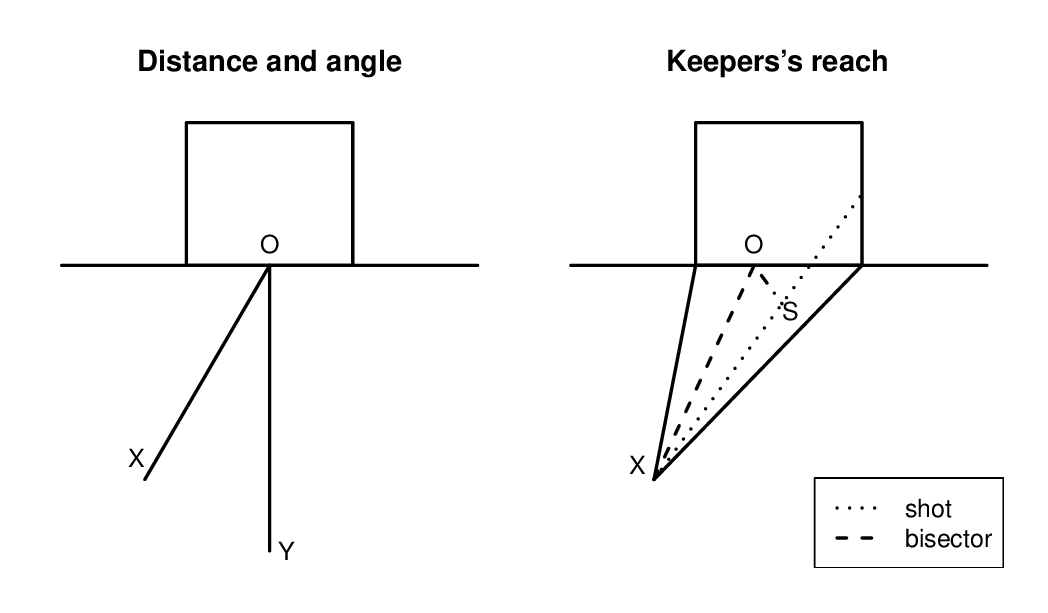}
\end{center}
\caption{(Left) For location X, the distance is the Euclidean distance between X and the center (O) of the goalline, while the angle (with appropriate sign) is $\angle XOY$, OY being the line that bisects the field horizontally. (Right) O is the ideal position of the goalkeeper (bisector of the angle the location of the shot makes with the goal), the line with small dots is the shot and OS is the keepreach, the shortest distance between the ideal location and the path of the shot.}
\label{fig:shot-location}
\end{figure}

Then, in order to fit an appropriate model, it is crucial to identify possible factors affecting the conversion of a shot. For that, the first thing we considered are the distance and angle of the location from where a shot is taken. The way we evaluated these two covariates is displayed on the left panel of \Cref{fig:shot-location}. It should be noted that this definition of the angle and the distance of a shot helps us consider them as uncorrelated covariates in this study. The right panel of the same figure shows another important covariate in our study, the keeper's reach. It is evaluated as the shortest distance the goalkeeper has to cover if he stands at the best position corresponding to the location of the shot.

\begin{figure}[!hbt]
\begin{center}
\includegraphics[width=\textwidth,keepaspectratio]{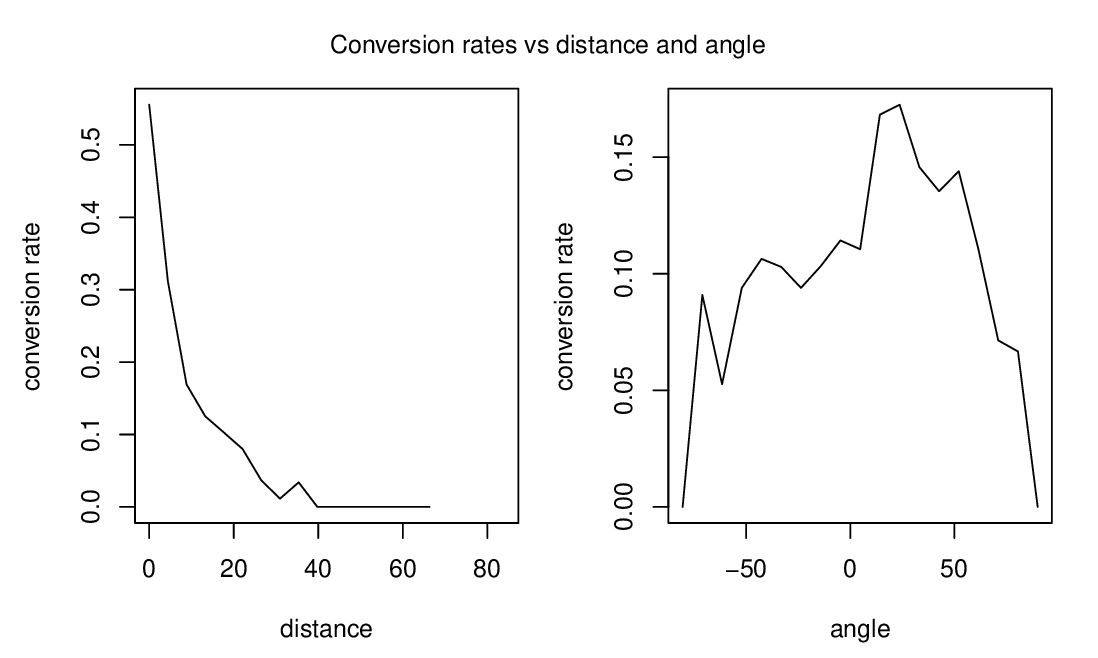}
\end{center}
\caption{Conversion rates corresponding to different distances (in yards) and angles (in degrees).}
\label{fig:rate-distance-angle}
\end{figure}

Further, exploratory analysis suggested that a transformation of variable is needed for the distance and the angle of a shot. Similar to the study by \citet{asa16}, we found that it is more appropriate to model the conversion rates as linear function of the logarithm of distance and cosine of angles instead of the usual values. The plots in \Cref{fig:rate-distance-angle} support this.

In addition to the distance, angle and keeper's reach, there are information on several other covariates and five of them are considered in this study - if it is a home game, the half of play when the shot is taken, body part used in the shot (header, left foot, right foot or other), goal difference at the time of the shot i.e. whether the shooter's team is leading, level or trailing, if the shot takes place during the stoppage time and the proportion of shots converted against the same opponent so far. 

Note that the set of covariates we are using includes several situational covariates (home or away, half of the game, goal difference, and the stoppage time) while the last covariate takes into account the strength of the opposing teams and helps us to identify if it significantly affects the rate of conversion. 

A quick summary of the total number and the proportions of shots converted for different levels of the aforementioned five factored covariates in the study are presented in \Cref{tab:covariates}. For example, the first row says that 2232 (56.4\% of the total of 3957) shots were taken by the home side and 286 of them (which amounts to 12.2\% of the shots taken by the home side) were converted successfully. 

One can clearly see that the home factor shows only a little change from one level to another. Same is true for first half versus second half. Very few shots were taken by other means than left or right foot or headers and it should not show significant effect in the analysis. For the three different types of body parts, the header effect is slightly less than the other two. As expected, the conversion rate of both the right-footed and left-footed shots are almost same, at least empirically, which accounts for the symmetry of the play. Finally, the conversion rate seems to be slightly higher when the shooter's team is leading while for the stoppage time shots, the conversion rate is 4\% more than the same for regulation time scenarios.

Further taking a deeper look into the covariate corresponding to body-parts, we noticed that the headers were usually taken from a close range, mostly from inside the box. In fact, the median distance from where the headers were taken was 11.2 yards, starkly different from the overall median of 17.6 yards. \Cref{tab:header-other} shows a summary of the distances and angles (in radian) for headers and other shots. The angles are similar for both cases, but the headers are always taken from a closer range. Naturally, the dependence on the covariates might be totally different for headers and other shots, and so, we suggest separate analysis for the two cases. It is worth mention that this has been overlooked in earlier studies.

\section{Motivation of spatially correlated error process}
\label{sec:motivation}

In most of the works discussed before, the usual practice is to fit a logistic regression with appropriate covariates. However, for these models, what we have commonly observed is a lack of goodness-of-fit checks, a moderately high average error rate and the use of wrong resorts to validate a model (e.g. usual R-squared in case of logistic regression). Even for the study by \citet{lucey14} - one of the most sophisticated techniques in our opinion - where conditional random field was used to take into account defender position and attacking context to model the probability of scoring a goal, we found the error rate to be 14.3\%, only 2\% less than what would have been for the usual logistic regression model without even considering defender attributes. And thus, there is a dire need to investigate these models in detail, so as to identify why they fail in practice. In this regard, the reader is instructed to read the \emph{Deadspin} article by \citet{bertin15}, who discussed the inefficiencies of common statistical models in practice in this regard.

We start our analysis with typical binary regression model with probit links for the MLS data we have. For compactness of the paper, we will discuss only the summary of what we observed with the typical model. First of all, for other shots (not headers), only the distance and angle, transformed to logarithmic and cosine scales, respectively, appeared to be significant. It was found that a smaller distance and a bigger angle significantly (both $p$-values very close to 0) improve the conversion rates. The estimate for the {\it keepreach} variable was estimated to be negative with a $p$-value of 0.067 while the same was positive with a $p$-value of 0.097 for the parameter corresponding to the factor when the shooter's team is leading. This shows that the conversion rate somewhat significantly increases when {\it keepreach} is less and when the shooter's team has a positive goal difference. The other covariates in the study showed no significant effect at all. 

On the other hand, for headers, it was found that a smaller distance, smaller keepreach and a higher angle increase the conversion chance significantly (all $p$-values less than 0.05), while the other covariates are not significant.

However, standard model diagnostics techniques revealed that the above models suffer from high deviance and high classification errors, when exposed to cross-validation. And so, we felt that there is a need to improve upon these models to build a feasible one. In this regard, our initial guess was that two shots may not at all be spatially independent and hence, we cannot take independence assumptions for model fitting. 

\begin{figure}[!thb]
\begin{center}
\includegraphics[width=\textwidth]{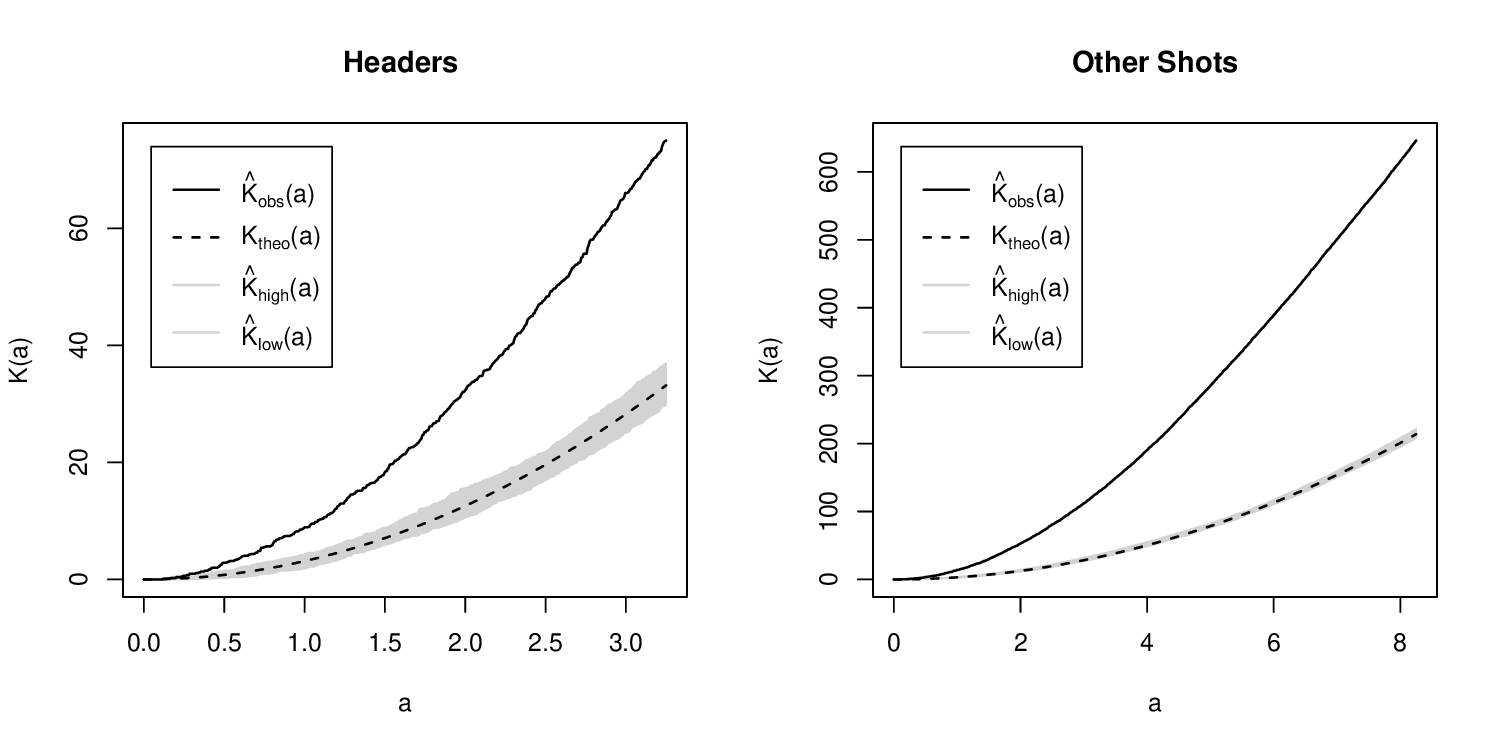}
\end{center}
\caption{Ripley's K-function $K(a)$ is plotted against $a$ for both headed and non-headed goals considering goals as spatial point processes. Under spatial homogeneity and independence, $K(a)=\pi a^2$. Expected values under this hypothesis and confidence intervals are constructed for reference.}
\label{fig:ripley}
\end{figure}

Recall that we can consider the occurrence of goals to be a spatial point process with the field being the location set. Plotting Ripley's K-function (\citet{dixon02}) is a standard technique to check whether there exists any spatial clustering or not. It is defined as $K(a)=\lambda^{-1}E_a$ where $\lambda$ is the density (number per unit area) of events (here goals), and $E_a$ is the expected number of extra events within distance $a$ of a randomly chosen event. In \Cref{fig:ripley}, $K_{theo}(a)$, $\hat K_{high}(a)$ and $\hat K_{low}(a)$ give us the expected values and confidence intervals of $K(a)$ under the assumption of spatial homogeneity and independence. We can clearly observe that the $K(a)$ values diverge from the expected values indicating possible spatial correlation and more spatial clustering as $a$ increases.

Now that we have a hint that there might be a spatial dependency involved, we test whether there exists spatial autocorrelation or not. We have a binary variable with $1$ denoting a goal and $0$ otherwise. We create a graph of the shot locations based on $k$-nearest neighbor criteria using Euclidean distance (taking $k=63$ as it has been accepted as a thumb rule to take $k \approx \sqrt{n}$ where $n$ is the number of observations). We join edges between two points if a point belongs to the set of $63$ nearest neighbours of another point and vice-versa. Thus, joining the neighbours, we can get the join-count statistics (\citet{cliff81}) to test for spatial autocorrelation. We perform join-count test against the alternative that the number of like joins are more than expected from random under 5\% significance level and found clear evidence for positive spatial auto-correlation as the number of $0-1$ join-counts were significantly lower than the expected values and both $1-1$ and $0-0$ counts were significantly higher than the expected ($p$-value nearly 0 for all the cases).

Note that positive autocorrelation indicates clustering of similar values and hence we conclude that there is sufficient spatial correlation to account for with the correlation between probabilities of shot conversions from two locations on the field decreasing  with increasing distance between the locations. Thus, one cannot perform the analysis considering spatial independence of the shots. Driven by the exploratory findings above, our assumption is that the shots are dependent even if they are from different matches and different players. This can be envisioned as a big shooting experiment in soccer which has been done on the same hypothetical field under different circumstances. We further emphasize that, to the best of our knowledge, this interesting and very important phenomena of spatial dependence has never been accounted for before in soccer. To tackle this, we introduce a spatially correlated error process in the standard probit model with the spatial covariance following an exponential decay law. 

On the other hand, in most of the real circumstances, the outcome of a shot significantly depends on the ability of the shooter, and hence this should be incorporated in the model. In order to quantify this, we include a random effect component, the posterior mean of which will evaluate the effect of human intervention in a shot, and that is what we term as `shooting prowess' or `heading prowess' in later sections. 

A detailed description of the model is provided in the next section. We use complete Bayesian framework to estimate the parameters and variances in the  model. This idea, in form of spatial autoregressive models, is close to what has been used in spatial econometrics before,  cf. \citet{elhorst2017transitions} and \citet{baltagi2016spatial}.

\section{Methods}
\label{sec:methods}

\subsection{Proposed model}
\label{subsec:proposed-model}

We start with a probit regression model equipped with a spatially correlated error process, in light of the discussion in the previous section. For shots $i=1,\hdots,N$, suppose $Y_i$ is the binary outcome (which takes the value 1 if the $i$th shot was converted to a goal), $s_i$ denotes the location of the shot and $X_i$ denotes the column vector with the values of the covariates. 

Our proposed model starts with the assumption that we have a latent variable $r_i$, such that $Y_i \sim \ber(p_i)$, $p_i=P(r_i>0)$. $r_i$ depends on the values of the covariates following a simple linear regression model, and $z_{m(i)}$ is an additive random effect that quantifies the player's ability. Here, $m(i)$ corresponds to the player who took the $i$th shot. In order to avoid the issue of over-fitting, we separately consider the players whose number of shots is more than a minimum cutoff, say $s_m$. Throughout this study, we use $s_m=10$. For all other players, ${m(i)}$ is considered to be same. If we use $Pl_1,\hdots,Pl_{M-1}$ to denote the players whose number of shots is greater than $s_m$, and we associate $z_j$ to $Pl_j$, then
\begin{equation}
    \label{eqn:player-id}
    m(i) = \begin{cases}
           j & \text{if $i$th shot is taken by $Pl_j$} \\
           M & \text{if $i$th shot is not taken by $Pl_1,\hdots,Pl_{M-1}$.} 
           \end{cases}
\end{equation}

We further assume that, given the values of the covariates, the outcomes will be dependent of each other through the shot locations and this dependence would decrease with distance. These assumptions support our findings while exploring the data in the previous section. 

We are going to consider a hierarchical structure. First of all, $Y_i=\mathbb{I}(r_i>0)$ and
\begin{equation}
\label{eqn:hierarchical-model}
r_i = X_i\dash \theta + z_{m(i)} + \eps_i,
\end{equation}
where $\theta$ is a parameter vector of appropriate order corresponding to the covariates, $z_{m(i)}$ is the player-effect on the shot, and $\eps_i$ is the error process. The term $X_i\dash \theta$ is considered to be an additive combination of the effects of some continuous and some discrete covariates. $z_{m(i)}$'s are considered to be independent and identically distributed (iid) random effects, coming from $N(0,\sigma_p^2)$ distribution. Note that there will be $M$ different random effects, if there are $M-1$ different players whose number of shots is greater than $s_m$. We further assume that the error process has the following structure:
\begin{equation}
\eps_i = w_i + e_i.
\end{equation}

Here $w_i$ denotes a zero-mean spatially correlated process and $e_i$ stands for a zero-mean independent and identically distributed (iid) white noise process. We assume that $e_i$'s are iid $N(0,\sigma^2)$ while for $w_i$'s, we take a correlation structure that decays exponentially. In particular, the covariance between $w_i$ and $w_j$ (i.e. for locations $s_i$ and $s_j$) is 
\begin{equation}
\label{eqn:spatial-covariance}
\cov(w_i,w_j) = \sigma_w^2 \ \exp(-\phi\norm{s_i-s_j}), 
\end{equation}
where the distance function $\norm{s_i-s_j}$ is taken as the Euclidean distance of the two locations.

Now, let us use $Y$ to denote the vector of all outcomes, $r$ and $p$ to denote the vector of $r_i$'s and $p_i$'s, respectively. $X$ is the design matrix such that $X\dash =[X_1:\hdots:X_N]$. We will denote the covariance matrix of $w$ by $\sigma_w^2\Sigma_w$. Finally, let $z$ be a vector of length $M$, corresponding to the player-effects. Then, the vector of $z_{m(i)}$'s can be written in the form $Az$, where each row of $A$ (a $N\times M$ matrix) denotes the index of the player effect. Thus, we can write the full model in the following form:
\begin{eqnarray}
\label{eqn:full-model}
Y_i &=& \mathbb{I}(r_i>0), \\ 
r &=& X \theta + Az + w + e, \nonumber \\
\text{such that } \ z &\sim& N(0,\sigma_p^2I_{M\times M}), \nonumber \\
w &\sim& N(0,\sigma_w^2\Sigma_w), \nonumber \\
\text{and } \ e &\sim& N(0,\sigma^2I_{N\times N}). \nonumber
\end{eqnarray}

We are going to use a Bayesian framework to get estimates for the parameters in the model. For each of the components in the parameter vector $\theta$, we will consider improper Jeffrey's prior distribution. On the other hand, we will assume the two variance components $\sigma^2,\sigma_w^2$ to be equal. While we acknowledge the simplicity associated with this assumption, it is worth mentioning that we have performed some simulation studies and found that the performance is approximately the same with the assumption. 
The computational complexity becomes cumbersome already with complicated priors in Gibbs sampling and the computational burden decreases manifold if we consider the variances equal. Also, we achieve satisfactory prediction power with this parsimonious assumption and hence, don't want to increase our number of parameters. Further discussion on this assumption is provided at the end of \Cref{subsec:gibbs}.

The priors we associate with $\sigma^2$ (same as $\sigma_w^2$) and $\sigma_p^2$ are independent inverse gamma (IG) distribution with parameters $a,b$. We would take $a>1$ to avoid improper posterior distributions.

And finally, the parameter $\phi$ (refer to \Cref{eqn:spatial-covariance}) will be kept fixed throughout the analysis. In order to find out the best possible value for $\phi$, we would consider a cross-validation scheme. The possible choices for $\phi$ used in the study were between 0.05 and 1. Based on the relationship $e^{-\phi d}\approx 0.05$, we chose the above range, which corresponds to actual distances in the field ranging from 3 yards to 60 yards. The validation scheme would consider prediction for $20\%$ of all the shots after estimating the parameters from the other $80\%$ and would find out the mean squared error for those predictions. For shots considered for the validation purpose (let us denote them by $i \in V=\{v_1,\hdots,v_m\}$), we would predict the probability of them being converted to goals and let the predicted probabilities be denoted by $\hat p_i$ while $Y_i$'s are the actual outcomes. Then, for each $\phi$, we calculate the mean squared error of the predictions using the following formula:
\begin{equation}
\label{eqn:validation-mse}
\text{MSE}_\phi = \frac{1}{m}\sum_{i\in V} (Y_i - \hat p_i)^2.
\end{equation}

The value of $\phi$ with the least mean squared validation error will be our optimal choice for the analysis. The prediction procedure is discussed in detail in \Cref{subsec:prediction}.

\subsection{Posterior distribution and Gibbs sampling}
\label{subsec:gibbs}

Throughout the discussion below, $K$ will always indicate a constant term, that may vary from time to time. Using the full model described in \Cref{eqn:full-model} and the prior distributions described in the paragraph following that equation, we can write the joint posterior distribution as:
\begin{eqnarray}
\log \pi(r,\theta,\sigma^2,w,z\given Y) &=& K + \sum_{i=1}^N\{Y_i\log(P(r_i>0))+(1-Y_i)\log(P(r_i\leq0))\} - \frac{w\dash \Sigma_{w}^{-1}w}{2\sigma^2}  \nonumber \\
\label{eqn:joint-posterior}  
&& - \frac{\norm{r-X\theta-Az-w}^2}{2\sigma^2}  - (a+N+1)\log \sigma^2 - \frac{b}{\sigma^2} \nonumber \\
&&  - \frac{\norm{z}^2}{2\sigma_p^2} - \left(a+\frac{M}{2}+1\right) \log {\sigma_p^2} - \frac{b}{\sigma_p^2}. 
\end{eqnarray}

Since simulating directly from the joint posterior distribution is difficult, we would use the principles of Gibbs sampling, which sequentially updates every parameter in an iterative way until convergence. For $\sigma^2$, we can get the full conditional distribution as:
\begin{equation*}
\log \pi(\sigma^2\given Y,r,\theta,w,z) = K - (a+N+1)\log \sigma^2 - \frac{1}{\sigma^2}\biggl[b + \frac12\norm{r-X\theta-Az-w}^2 + \frac12w\dash \Sigma_{w}^{-1}w\biggr] 
\end{equation*}
and thus, 
\begin{equation}
\label{eqn:sigma-posterior}
(\sigma^2\given Y,r,\theta,w,z) \sim IG\left(a+N,b+\frac12\norm{r-X\theta-Az-w}^2 + \frac12w\dash \Sigma_{w}^{-1}w\right).
\end{equation}

Similarly, we get that
\begin{equation}
\label{eqn:sigmap-posterior}
(\sigma_p^2\given Y,r,\theta,w,z) \sim IG\left(a+\frac M2,b+\frac{1}{2}\sum_{k=1}^{m}{z_k^2}\right).
\end{equation}

For $\theta$, we get that
\begin{equation*}
\log \pi(\theta \given Y,r,\sigma^2,w,z) = K - \frac12 \theta\dash \biggl(\frac{X\dash X}{\sigma^2}\biggr)\theta + \frac12 \cdot 2\theta\dash  \cdot \frac{X\dash (r-Az-w)}{\sigma^2}.
\end{equation*}
Straightforward calculations then tell us that the conditional posterior of $\theta$ is 
\begin{equation}
\label{eqn:theta-posterior}
(\theta \given Y,r,\sigma^2,w,z) \sim N\left((X\dash X)^{-1}X\dash (r-Az-w),\sigma^2(X\dash X)^{-1}\right).
\end{equation}

On the other hand, using similar techniques, we can get the posterior distribution for $w$ as:
\begin{equation}
\label{eqn:w-posterior}
(w \given Y,r,\theta,\sigma^2,z) \sim N\left((I+\Sigma_{w}^{-1})^{-1}(r-X\theta-Az),\sigma^2(I+\Sigma_{w}^{-1})^{-1}\right).\end{equation}

Further, if we define $\sum_{i=1}^N{\mathbb{I}\{m(i)=k\}}=n_k$, we can get the posterior distribution for $z_k$ as:
\begin{equation}
\label{eqn:z-posterior}
(z_k \given Y,r,\theta,\sigma^2,w) \sim N\left(\left(n_k+\frac{\sigma^2}{{\sigma_p}^2}\right)^{-1}\sum_{i:m(i)=k}{(r_i-X_i\dash\theta-w_i)},\left(\frac{n_k}{\sigma^2}+ \frac{1}{{\sigma_p}^2}\right)^{-1}\right).
\end{equation}

Note that  the means of the conditional distributions of both $\theta$ and $w$ require the computation of the products of a matrix and a vector, but the dispersion matrices of the same are updated at every step only through the value of $\sigma$ and so, one needs to compute the inverses of three big matrices, namely $X\dash X$, $\Sigma_w$ and $(I+\Sigma_w^{-1})$, only once throughout the MCMC analysis. Hence, directly using the above conditional distributions at every step will not suffer from huge computational burden. However, one can always use component-wise conditional posterior distributions to further increase the efficiency.

Finally, we find the conditional distribution of $r$. The joint posterior distributions above clearly show that conditional on data and other parameters, each component of $r$ is independent of others and simple calculations would reveal that the conditional distributions are truncated normal as follows.
\begin{equation}
\label{eqn:r-posterior}
(r_i\given Y,\theta,w,\sigma^2,z) \sim \begin{cases} TN(\alpha_i,\sigma^2;0,\infty) & \text{if } \ Y_i=1 \\ TN(\alpha_i,\sigma^2;-\infty,0) & \text{if } \ Y_i=0 \end{cases}
\end{equation}
In the above, $\alpha_i$ is $i$th component of $\alpha=X\theta+w+Az$ and $TN(\mu,\tau^2;a,b)$ denotes a normal distribution with mean $\mu$ and variance $\tau^2$, truncated in the interval $[a,b]$.

Now to note that, if we assume $\sigma_w^2 \ne \sigma^2$, then the posterior distribution of $w$ (see \Cref{eqn:w-posterior}) will become
\begin{equation}
\label{eqn:w-posterior-revise}
(w \given Y,r,\theta,\sigma^2,\sigma_w^2,z) \sim N\left(\sigma^2(I+\frac{\sigma^2}{\sigma_w^2}\Sigma_{w}^{-1})^{-1}(r-X\theta-Az),\sigma^2(I+\frac{\sigma^2}{\sigma_w^2}\Sigma_{w}^{-1})^{-1}\right).\end{equation}

This implies that at each step of Gibbs sampling, we will have to invert a $n\times n$ matrix $(I+\frac{\sigma^2}{\sigma_w^2}\Sigma_{w}^{-1})$, which requires $O(n^3)$ time. Naturally, it is computationally not feasible to do it for thousands of iterations when $n$ is large (in our case $n=3957$). Whereas, if we follow $\Cref{eqn:w-posterior}$, we will have to invert the matrix $(I+\Sigma_{w}^{-1})$ only once and we can use that in subsequent iterations. Hence, this parsimonious assumption leads to a huge decrease in computational complexity and we do not lose much in prediction. 

\subsection{Future prediction}
\label{subsec:prediction}

In order to make a new prediction for a shot from location $s\dash$, we will take resort to the posterior estimates obtained from the Gibbs sampler. Let us denote the outcome of the shot from $s\dash$ by $Y(s\dash)$ ($X(s\dash)$, $r(s\dash)$, $z(s\dash)$ and $w(s\dash)$ are defined accordingly). Note that the posterior predictive distribution $f(Y(s\dash)\given Y)$ can be written using an integral for the product of the density functions, where the integral is taken with respect to all the parameter vectors. However, instead of solving that integral, a better and more convenient idea is to use the Gibbs sampler estimates to draw observations from the posterior predictive distribution and one can do it sequentially.

At first, we draw samples for $r,w,\theta,\sigma^2$ using the conditional posterior distributions described in \Cref{subsec:gibbs}. Also, if it comes from $k$th player used in our training data, we define $z(s\dash)$ as the posterior mean of $z_k$, else, we simulate $z(s\dash)$ from $N(0,\sigma_p^2)$. After that, we draw samples for $w(s\dash)$ using the conditional distribution of $(w(s\dash)\given w,\sigma^2)$ as discussed below. Using the generated observations, an estimate for $r(s\dash)$ can be obtained by taking a realization from a normal distribution with mean $X(s\dash)\dash \theta+w(s\dash) + z(s\dash)$ and variance $\sigma^2$. Finally, we set $Y(s\dash)=\ind(r(s\dash)>0)$. Alternately, one can also use $Y(s\dash) \sim \ber(p(s\dash))$, where $p(s\dash)=P(r(s\dash)>0)$ can be computed using the sampled values of $\theta$, $\sigma^2$, $z(s\dash)$ and $w(s\dash)$.

To compute the conditional distribution of  $(w(s\dash)\given w,\sigma^2)$, we start with the fact that
$$\left(\begin{array}{c} w(s\dash) \\ w\end{array}\right) \sim N\biggl(0,\sigma^2\left[\begin{array}{cc} 1 & \Sigma_{s-s\dash}\dash  \\ \Sigma_{s-s\dash} & \Sigma_{w} \end{array}\right]\biggr).$$

Here, $\Sigma_{s-s\dash}$ is the column vector denoting the covariance of $w(s\dash)$ with the elements of $w$. Following \Cref{eqn:spatial-covariance}, the $j$th element of $\Sigma_{s-s\dash}$ is $\exp(-\phi\norm{s_j-s\dash})$. Using the principle of conditional distribution for multivariate normal distribution, we can then say that
\begin{eqnarray}
\label{eqn:wnew-conditional}
w(s\dash)\given w &\sim& N(\mu_{\text{new}},\sigma^2_{\text{new}}), \\ 
\text{where } \ \sigma_{\text{new}}^2 &=& \sigma^2\bigl(1-(\Sigma\dash _{s-s\dash}\Sigma_w^{-1}\Sigma_{s-s\dash})\bigr), \nonumber \\ 
\mu_{\text{new}} &=& \Sigma\dash _{s-s\dash}\Sigma_w^{-1}w. \nonumber
\end{eqnarray}

\subsection{Measures to quantify player abilities}
\label{subsec:measures}

Based on our model, we propose two measures to quantify a player's ability on the pitch. 


To start with, we would like to introduce, what we call, the {\it Shooting Prowess (SP)} of a player. We define this measure ${SP}_k$ as the posterior mean of $z_k$ i.e. the player-specific random effect. Recall that we use a cutoff of $s_m=10$ matches to evaluate this random effect, and hence, discrete cases of players taking few shots to score will be considered as general effect in our model. This would effectively help us in identifying the players who are in general good shooters.

Next, we would like to quantify a player's {\it Positioning sense (PS)}, which is a pivotal thing for a striker on the field in terms of scoring goals. One has to be at the right place at the right time to get a goal in his name. From our training model, we can estimate the shot conversion probability of an average player from a particular distance, angle and type of opportunity. It is a different issue whether the player has grabbed the opportunity or not, but getting into a perfect position, he could ensure that a shot was fired. 

In order to do that for the $k$th player, suppose that in a total of $g_k$ matches,  he has taken $n_k$ shots (calculated over a time period, possibly over a whole season or for the entire career) in total. The expected probability of the $i$th shot, $i=1,\hdots,n_k$, is denoted by $\hat p_i$. Then, summing up the predicted probability of shot conversion for that player can give us a perfect idea about his positioning sense. Necessarily, the ${PS}_k$ (Positioning Sense) of the $k$th player is defined by
\begin{equation}
\label{eqn:ps}
{PS}_k=\frac{1}{g_k}\sum_{i=1}^{n_k} {\hat p_i}.
\end{equation}

The fact that we calculate the average over the number of matches and not over the set of all shots captures the information of whether a player takes more shots in general.

\section{Results}
\label{sec:results}

\subsection{Fitting the model}

Following the discussions in the previous sections, we consider two different models for `other shots' and `headers', and they can be written as follows.

For other shots, let $Y_i=\ind(r_i>0)$ denote whether $i$th shot was a goal. Then, we set $r_i=\mu_i+z_{m(i)}+w_i+\epsilon_i$, where $z_{m(i)},w_i,\epsilon_i$ are as described in the previous section and $\mu_i$ is the mean function defined by
\begin{eqnarray}
\mu_i &=& \beta_0 + \beta_1\log(D_i) + \beta_2 \cos(A_i) + \beta_3 K_i + \beta_4 P_i + \beta_5\ind\{H_i=1\} + \beta_6\ind\{F_i=1\}  \nonumber \\
\label{eqn:model-mean}
&&  + \sum_{j=0}^1\beta_{7,j}\ind\{GD_i=j\} + \beta_8\ind\{S_i=1\} + \sum_{k=2}^3 \beta_{9,j}\ind\{BP_i=j\}.
\end{eqnarray}

Here, $D_i$, $A_i$, $K_i$ and $P_i$ denote the continuous covariates, namely distance, angle, keepreach and proportion of shots converted against the same opponents. Indicators for $H_i$, $F_i$ and $S_i$ equal to 1 are used to denote if it was a home game, if it was in the first half and whether it happened during stoppage time. $GD_i$ denotes whether the goal difference at the time of the shot was zero ($j=0$) or positive ($j=1$). And finally, $BP_i$ shows which body part ($j=1,2,3$ corresponding to other, left foot and right foot) was used in taking the shot. Note that we take $\beta_{7,-1}=\beta_{9,1}=0$ for identifiability purpose.

On the other hand, for `headers', we do not consider the covariate $BP$, but the other model remains unchanged.

Now, using the mean function (\Cref{eqn:model-mean}), we fit the model described in \Cref{sec:methods}. First of all, using the procedure described towards the end of \Cref{subsec:proposed-model}, we evaluate the best choice of $\phi$ to be used in the spatial covariance function (\Cref{eqn:spatial-covariance}). It is obtained to be 0.75 (and 0.45) for `headers' (and for `other shots). Following the approximate relation $e^{-\phi d}\approx 0.05$, we can say that the spatial correlation in converting headers (and other shots) is insignificant for locations approximately 4 yards (and 6.7 yards) or more apart from each other. Then, we use the Gibbs sampler to estimate the parameters. We have used 10000 runs in the burn-in period and have used MCMC samples of size 1000 to evaluate the estimates, standard errors and the confidence intervals of the parameters. Summary of the results for the `headers' are shown in \Cref{tab:estimates-headers}, while the same for `other shots' are displayed in \Cref{tab:estimates-others}.

From the results, we can see that the effect of logarithm of distance is significantly negative for both headers and other shots. This matches with the established theory that the closer you shoot from, the greater chances of the shot being converted to goal. However, the conclusion about the `angle' of shots is interesting. In the simple logistic regression model used by current researchers in this field, it was found that a smaller angle (higher value of the cosine of the angles) significantly increases the goal-scoring chance of any shot. But, in our model, we have found that the angle of the shot is not a significant covariate for headers. This happens probably because the effect of the angle is already being explained by the spatial process. On the contrary, even with the spatial dependence in the model, the angle has significant positive effect on the conversion rate for other shots.

So far as the other variables are concerned, only keeper's reach and stoppage time are significant for headers. The chance of converting from a header increases when the keeper's reach is less. Also, headers taken during the stoppage time has a significantly higher chance of being converted. On the other hand, we also find that the two variance components are very different. While the spatially correlated process shows a variance of 0.999, the player effect explains more variability (estimated value of $\sigma_p^2$ is 1.656) in the data.

Further, for non-headed shots, barring the distance and angle of the shot, no other situational or match-specific covariate is significant. One interesting observation is the estimates and confidence intervals for the effect of left or right footed shots. They are very close to each other, thereby supporting the nature of symmetry in the game of soccer. The variability in the data, similar to the previous case, is explained more by the player-effect ($\sigma_p^2$ is estimated to be 1.675) while the spatially correlated process has a variance of 0.995. These values are similar to the `headers' data.

\subsection{Diagnostics}

We next delve into the diagnostics of the model, and compare the performance of our model to three different models. First, we consider the simple logistic regression model (SLRM) which is the most common one to use for such data. Second, we consider the spatial probit model proposed by \citet{klier08}. It is in essence similar to our proposed model. This model, henceforth denoted as KMM, uses a spatially related weight matrix, where the weights are corresponding to the number of observations contiguous to every observation. The parameters, including the spatial interaction coefficient, are estimated by a linearized version of generalized method of moments estimators for probit models. We include this model in this comparison study to understand if our specification of the spatial correlation can be deemed appropriate. Finally, we include a machine-learning algorithm. A type of neural network (later denoted as NN) that performs well on such a binary classification problem from a set of vectors is a simple stack of fully connected (“dense”) layers with rectified linear unit activation function. We use two dense layers with $16$ and $8$ hidden units and then pass it onto a layer with sigmoid function as output to train classification probabilities.

In order to compare the performance of the different techniques, at first we calculate the Brier score for all the models. This is recognized as one of the most important score function to measure the accuracy of model fit for binary type data. It is defined as $\sum_{t=1}^{N}{(f_t-o_t)}^2/N$ where $f_t$ is the expected probability of forecast and $o_t$ is the actual outcome. For the non-headed shots, we compute it to be 0.067, while for the headers it is approximately 0.061. Note that for both cases, this score is around 30\% less than what we observed in the other three candidate models, namely SLRM, KMM or NN. 

Another standard method to evaluate probabilistic prediction is the log score, which is defined by $\sum_{t=1}^N\{o_t\log f_t+(1-o_t)\log(1-f_t)\}$. A higher value of the log score will indicate a better predictive ability of the model. We have observed improvement in our model according to this measure as well. For headers, SLRM, KMM and NN yield log score of approximately $-301.7$, $-338.9$ and $-299.4$, respectively, while the same for our model is nearly $-184.3$. Thus, our model performs more than $38\%$ better in this aspect. Similar phenomena has been observed for non-headed shots as well.

Next, we evaluate the proportion of times the models could classify the outcomes correctly. Here, the improvement with our model is about 20\%, as compared to the other three methods.

The above results are summarized in table \ref{tab:comparison}.

As a final piece, we want to see how accurately we can predict out-of-sample outcomes based on the model we developed. At this point, one should keep in mind that this is a case of rare events binary regression and one of the main issues with these scenarios is that the prediction problem is immensely difficult and some calibration method is always needed, which further requires more information on the whole population, as pointed out by \citet{king01}. Since we do not have that information, we have to rely on usual prediction methods (refer to \Cref{subsec:prediction}). We use cross-validation techniques to evaluate how good the predictions are, by taking a subset of the full data (approximately 80\%) to learn the model and applying it on the rest (approximately 20\%). Afterwards, in order to evaluate the prediction efficiency, we follow the general scoring rules outlined by \citet{merkle13}. 

\begin{figure}[!hbt]
\begin{center}
\includegraphics[width=\textwidth]{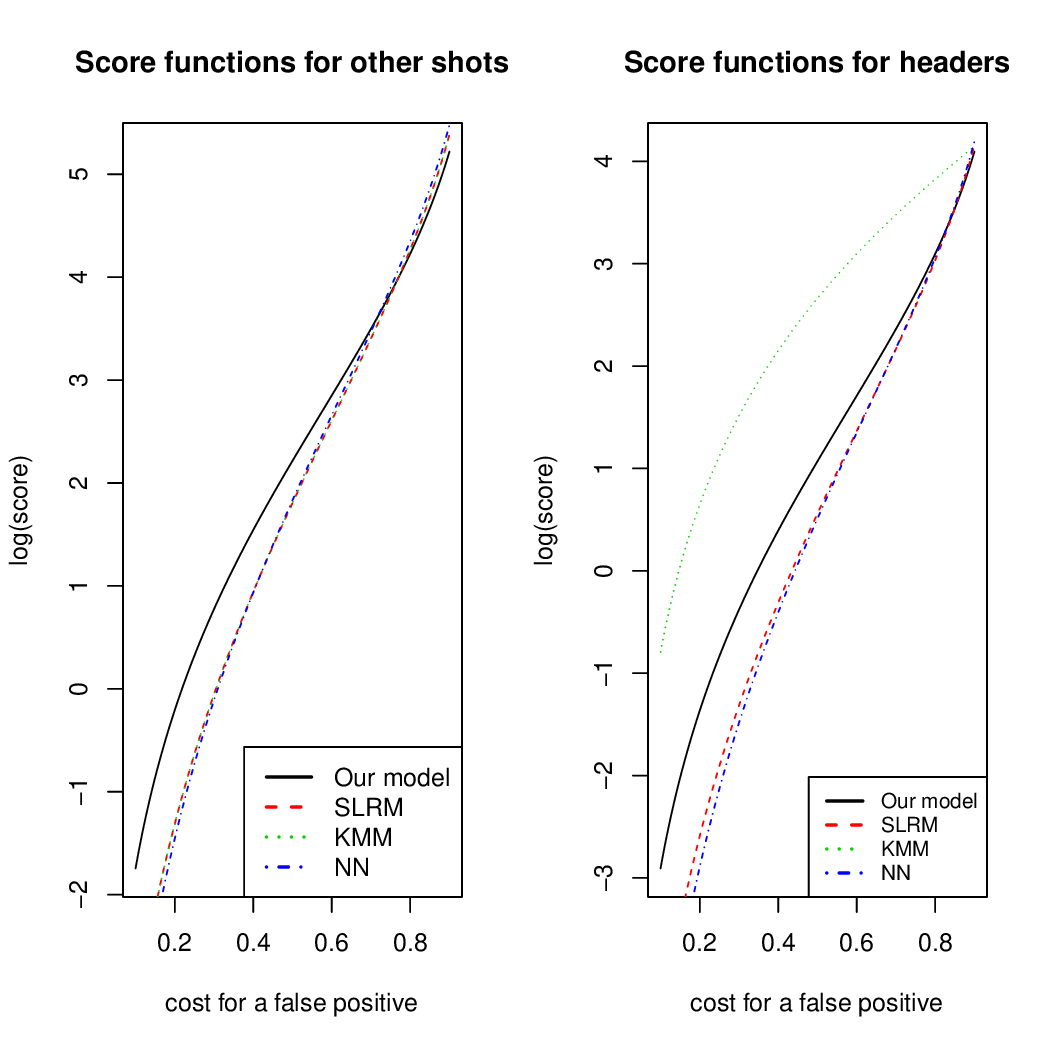}
\end{center}
\caption{Logarithm of the score function, corresponding to different cost values, for simple logistic regression model (SLRM), Klier-McMillen model (KMM), Neural network (NN) and for our proposed model.}
\label{fig:score-cost}
\end{figure}

In the aforementioned study, the authors discussed the beta family of proper scoring rules, proposed by \citet{buja05} for binary regression problems, in detail. Using different combinations of the parameters $(\alpha,\beta)$ for the beta distribution, one can choose the most appropriate scoring rule. For example, $\alpha=\beta=0$ corresponds to log score while $\alpha=\beta=1$ corresponds to Brier score. On the other hand, \citet{merkle13} presented another way of thinking about the parameters, where one can assume that the value of $\alpha/(\alpha+\beta)$ (or $1-\alpha/(\alpha+\beta)$) is the cost of a false positive (or false negative). In light of this, one should note that scoring rules with $\alpha<\beta$ emphasize low-probability forecasts, in the sense that such a scoring rule would heavily penalize a prediction that attach a low probability to a successful outcome (goals). Similarly, scoring rules with $\alpha>\beta$ emphasize high-probability forecasts. Using the R package {\sf scoring} (\citet{Scoring}), we have computed the scores for different values of $\alpha/(\alpha+\beta)$. The logarithm of the scores for the three models are plotted against different cost functions in \Cref{fig:score-cost}. 

We can see that the performances of the three methods are very similar for other shots. But, our model shows consistently better (higher score function) performance when the cost for a false positive is  less than 0.7. From a practical point of view, it is probably more important to predict a goal successfully, and naturally, the cost for a false positive should be less. And we can see that in such cases, our model performs better than the other two. Even for $\alpha=\beta$, when the cost is equal for goal and miss, calculations showed that for our model, the score function is approximately $20\%$ higher than the other candidate models.

For headers, though, the results are a bit different. While similar conclusions remain true for our model as compared to SLRM or NN, KMM performs the best. Hence, one can expect that for headers, the specification for the spatial correlation is probably not being best explained by the exponential decay function

To conclude this section, we once again emphasize that goals are example of rare events (\citet{anderson13}), as only about $10\%$ of the shots get converted to goals on an average. It is nearly impossible to get a very high prediction accuracy without calibration. Since we do not have the necessary information to calibrate the predictions, we should primarily focus on validating the model. And from the above discussions, we can say that in our proposed model performs better than the three different types of methods we considered.

\subsection{SP and PS of the players}
\label{subsec:sp-ps}

\begin{figure}[!hbt]
\begin{center}
\includegraphics[width=\textwidth]{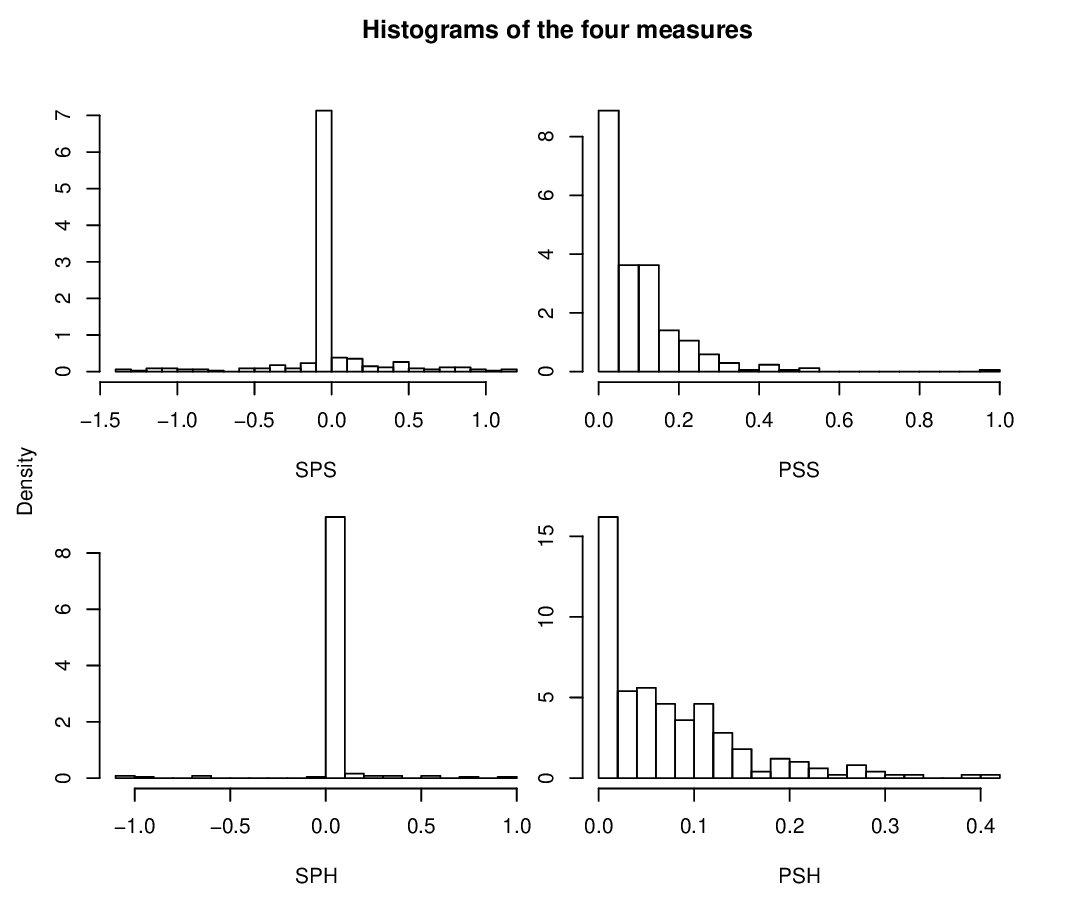}
\end{center}
\caption{Histogram of the shooting prowess (left) and positioning senses (right) for 381 players, calculated based on fitted probability and actual outcomes.}
\label{fig:histogram-sp-ps}
\end{figure}

We now proceed to compute the two measures described in \Cref{subsec:measures}. In line with the way we analyze the data, SP and PS are evaluated separately for headers (denoted as SPH and PSH henceforth) and other shots (denoted as SPS and PSS henceforth), in order to identify the players with best heading or shooting ability. Histograms of the measures for all these players are shown in \Cref{fig:histogram-sp-ps}. 

For the SP values in both cases (figures on the left), one can see that there is a huge peak around 0 and that the histogram is slightly skewed to the right, suggesting that with more shots per game, players become more adept in converting them to goals. Similarly for the figures on the right, we can see that most players showed standard positioning senses, while there were some outliers as well. 

Next, to examine if our measures can successfully capture the shooting abilities of the players, we identified the top five shooters in terms of SPS and PSS. However, at this point, to avoid potential small-sample issues, we considered players with more than 5 headers or more than 20 shots. The list of the best players in this regard, along with their real life data from the MLS 2016/17 season are presented in table \Cref{tab:sp-ps}. In that table, we also present the data for the top players according to SPH and PSH. 

We note that Ignacio Piatti, Giovani Dos Santos, Fanendo Adi, Ola Kamara, Bradley Wright-Phillips and Dominic Dwyer were in the list of top ten highest scorers in 2016/17 season and our measures show that the reason behind this is that all have great shooting or heading abilities. Michael Barrios, a winger for FC Dallas that season, is very good with his positioning senses and shooting abilities. Another winger, Kekuta Manneh also showed good shooting prowess and found a spot in the list of top five.  

The results for the heading abilities were something intriguing. At first, we like to point out the special cases of Drew Moor, Sebastian Hines and David Horst. All of them are defenders and their heading abilities are supposed to be good. In soccer, it is often observed that the defenders move ahead during a corner or a long free-kick. Naturally, it would be beneficial for the teams to identify which defenders can take good positions during the set-piece situations and convert the chances that come in the way. It is evident that our model can serve that purpose well. Among forwards, Bradley Wright-Phillips, Cyle Larin and Dominic Dwyer showed really good positioning sense and conversion ability when it comes to heading. Moreover, it was found that positioning sense is significantly positively correlated with heading prowess.

It is worth mention here that the training data we use in our model contained information only from a small part of the season and the above measures are always calculated based on that data. Thus, it is interesting that the list of best players according to these measures consists mostly of the players who were actually at the top of the goal-scoring chart after the season ended. This clearly reflects that we have been able to single out the players' goal-scoring abilities and positioning senses perfectly. In real life, it has huge potential to identify most suitable players in different scenarios.

\section{Summary}
\label{sec:conclusion}

In conclusion, we have presented an alternate way of analyzing the conversion rate of shots in soccer. Using a spatially correlated error process, we have shown that our model fits the data well. Our proposed method has exhibited improved Brier score, log score and predictive abilities, as compared to the logistic regression model that is in practice at the moment. On the other hand, our way of specifying the spatial correlation term is usually more appropriate than other possible way for probit regression models. It also provided better predictive abilities than neural network, one popular machine learning technique. 

Another key contribution of our work is to properly quantify the shooters' efficiency and positioning sense. To the best of our knowledge, this is the first paper to introduce such measures. We have also established that the quantitative measures are good enough to identify the best strikers and thus, the teams would certainly be benefited from using this concept. In a more specialized way, conditioning on different opponents, one can use relevant data to find out the best shooters against them and that would help the coaches in making strategies and tactics against particular teams.

On a related note, we also mention that using a similar measure for the goalkeepers, one can quantify the `saving prowess' to single out the best custodians in the league. We chose not to do that in this study, for we think the goalkeepers' abilities should not be evaluated solely based on the number of goals conceded, but also on the number of blocks and saves made by the goalkeeper. And that is one of the future directions we are interested in. We emphasize that the methods we described here are applicable to binary data (goals and misses), but it can be extended to the multinomial case (blocked, goal, missed, hit by post or saved) as well and that is what needs to be done to identify the best goalkeepers in business. 

A minor issue with our model is that if we have the exact same co-ordinates for two shot locations, then the covariance matrix associated with the error process will be singular and the model will fail. But, theoretically, it is a measure zero event and it happens very rarely in practice. For example, we did not encounter such an issue in our data. To understand it a bit further, observe that penalty kicks are one such kind of events, where the shots are taken from a fixed location inside the box. But, penalty kicks should not be considered in a study like ours because the outcome of the spot-kicks do not really depend on the distance, angle or other physiological attributes. Rather, it only depends on the penalty taker's abilities and usually every team has a specialist for this job. Hence, quite rightly, we left out these cases from the data before running the analysis. Now, apart from penalty kicks, even if we have two or more shots fired from the same exact location, we can easily make adjustments as follows. Suppose, $z_1,..,z_k$ are the shots fired from exact same location. Then, we randomly pick one shot from $z_1,..,z_k$  and leave out the other shots from the data. Next, we fit the model based on the available data and predict the outcomes of the left-out shots from the model. We calculate the sum of squared prediction errors in this case and continue this procedure for all of the $k$ shots. And finally, we pick the shot with the lowest error and include it, discarding the rest, in our final model.

We also note that the parameter $\phi$ (refer to \Cref{eqn:spatial-covariance}), which denotes the extent of spatial dependence, is not estimated directly from our model. Instead, we use a cross-validation approach to find out the best choice of $\phi$ from a grid. Naturally, this does not allow us to test the hypothesis of no spatial dependence, cf. \citet{anselin2001rao}. While it might seem a disadvantage, we want to point out that this is a common practice in many spatial studies, see, for example, \citet{sahu06}. In fact, in this study, we built the model after getting sufficient evidence in favour of spatial autocorrelation and hence, afterwards, we are more interested in explaining the variation among conversion of shots explained by the covariates and the spatial error process and then, how accurately we can predict the outcomes. The results establish that our model works well to serve that purpose.

Finally, we believe that the predictions will improve with more intense data containing information like the shot speed, the attacking speed and the position of the defenders in the shooting range of the shooter. As our model is really flexible, in future, we aim to incorporate those attributes in the discussion to conduct a more comprehensive study.

\newpage

\section{Acknowledgement}
\label{sec:acknowledgement}

We thank Dr. Chris Anderson, the author of {\it The Numbers Game : Why Everything You Know About Soccer Is Wrong} and a pioneer in soccer analytics, for providing his valuable feedback about our approach to address the research problem. Our gratefulness also goes to American Soccer Analysis (ASA) for providing the data in this study.

\newpage

\bibliographystyle{agsm}
\bibliography{reference}

\newpage

\section{Tables}







\begin{table}[!hbt]
\caption{Conversion rates for different levels of the factored covariates: For goal or miss, proportions given in brackets are conversion rates for that level while the proportion in the last column is the proportion of shots taken for that level}
\centering
\label{tab:covariates}
\vspace{0.1in}
\begin{tabular}{lccc}
\hline
Factor & Goal  & Miss & Total  \\
\hline
Home & 286 (12.2\%) &  1946 (87.8\%) & 2232 (56.4\%) \\
Away & 196 (11.4\%) & 1529 (88.6\%) & 1725 (43.6\%) \\
\hline
First half & 209 (12.0\%) &  1536 (88.0\%) & 1745 (44.1\%) \\
Second half & 273 (12.3\%) & 1939 (87.7\%) & 2212 (55.9\%) \\
\hline
Header & 106 (11.0\%) & 855 (89.0\%) & 961 (24.3\%) \\
Left Foot & 130 (12.5\%) & 911 (87.5\%) & 1041 (26.3\%) \\
Right Foot & 242 (12.5\%) & 1698 (87.5\%) & 1940 (49.0\%) \\
Other & 4 (26.7\%) & 11 (73.3\%) & 15 (0.4\%) \\
\hline
Leading & 127 (13.9\%) &  784 (86.1\%) & 911 (23.0\%) \\
Scores level & 218 (11.4\%) & 1687 (88.6\%) & 1905 (48.1\%) \\
Trailing & 137 (12.0\%) & 1004 (88.0\%) & 1141 (28.8\%) \\
\hline
Regulation time & 447 (11.9\%) & 3290 (88.1\%) & 3737 (94.4\%) \\
Stoppage time & 35 (15.9\%) & 185 (84.1\%) & 220 (5.6\%) \\
\hline
\end{tabular}
\end{table}

\newpage

\begin{table}[!hbt]
\caption{Comparison of shot locations (mean, median and standard deviation) for headers and other shots}
\centering
\label{tab:header-other}
\vspace{0.1in}
\begin{tabular}{lcccccc}
\hline
& \multicolumn{3}{c}{Summary of distance (in yards)} & \multicolumn{3}{c}{Summary of angle (in radian)} \\
\hline
        &   Mean  & Median & St. Dev. & Median & Mean & St. Dev. \\
\hline
Header      & 11.3 & 11.2 & 3.92 & $-0.007$ & 0.005 & 0.53 \\ 
\hline
Other shots & 22.0 & 20.8 & 8.77 & $0.004$ & 0.019 & 0.62 \\
\hline
\end{tabular}
\end{table}

\newpage

\begin{table}[!hbt]
\caption{Results from fitting our model to headers (total number of events was 961)}
\centering
\label{tab:estimates-headers}
\vspace{0.1in}
\begin{tabular}{lccc}
\hline
Parameter & Estimate & Standard error & Confidence interval \\
\hline
$\beta_0$ (Intercept)  		    & $0.349$ 	& 1.140 & $(-1.950,2.542)$ \\
$\beta_1$ (log(Distance)) 		& $-1.139$ 	& 0.389 & $(-1.904,-0.360)$ \\
$\beta_2$ (cos(Angle)) 			& $0.625$ 	& 0.874 & $(-1.062,2.336)$ \\
$\beta_3$ (Keeper's reach)  	& $-0.208$ 	& 0.058 & $(-0.326,-0.101)$ \\
$\beta_4$ (Opponent)     		& $-1.297$ 	& 1.625 & $(-4.324,1.863)$ \\
$\beta_5$ (Home)		 	    & $0.037$ 	& 0.154 & $(-0.262,0.337)$ \\
$\beta_6$ (First half) 		    & $0.002$ 	& 0.166 & $(-0.326,0.333)$ \\
$\beta_{7,0}$ (score level) 	& $0.052$ 	& 0.174 & $(-0.277,0.406)$ \\
$\beta_{7,1}$ (leading) 	    & $0.324$ 	& 0.211 & $(-0.078,0.747)$ \\
$\beta_8$ (Stoppage time)	    & $0.596$ 	& 0.292 & $(0.008,1.187)$ \\
$\sigma^2$ (Variance) 	    	& $0.999$ 	& 0.065 & $(0.884,1.129)$ \\
$\sigma_p^2$ (Player variance) 	& $1.656$ 	& 0.785 & $(0.787,3.687)$ \\
\hline
\end{tabular}
\end{table}

\newpage

\begin{table}[!hbt]
\caption{Results from fitting our model to other shots (total number of events was 2996)}
\centering
\label{tab:estimates-others}
\vspace{0.1in}
\begin{tabular}{lccc}
\hline
Parameter & Estimate & Standard error & Confidence interval \\
\hline
$\beta_0$ (Intercept)  		    & $1.506$ 	& 0.959 & $(-0.293,3.508)$ \\
$\beta_1$ (log(Distance)) 		& $-1.688$ 	& 0.253 & $(-2.208,-1.234)$ \\
$\beta_2$ (cos(Angle)) 			& $1.259$ 	& 0.604 & $(0.114,2.433)$ \\
$\beta_3$ (Keeper's reach)  	& $-0.055$ 	& 0.034 & $(-0.118, 0.011)$ \\
$\beta_4$ (Opponent)     		& $-0.320$ 	& 0.859 & $(-2.097,1.291)$ \\
$\beta_5$ (Home)		 	    & $0.012$ 	& 0.086 & $(-0.152,0.191)$ \\
$\beta_6$ (First half) 		    & $0.011$ 	& 0.089 & $(-0.167,0.189)$ \\
$\beta_{7,0}$ (score level) 	& $-0.030$ 	& 0.105 & $(-0.236,0.184)$ \\
$\beta_{7,1}$ (leading) 	    & $0.141$ 	& 0.121 & $(-0.083,0.366)$ \\
$\beta_8$ (Stoppage time)	    & $-0.058$ 	& 0.190 & $(-0.439,0.290)$ \\
$\beta_{9,2}$ (Left foot) 		& $0.553$ 	& 0.479 & $(-0.401,1.476)$ \\
$\beta_{9,3}$ (Right foot) 		& $0.613$ 	& 0.478 & $(-0.333,1.519)$ \\
$\sigma^2$ (Variance) 	    	& $0.995$ 	& 0.038 & $(0.925,1.073)$ \\
$\sigma_p^2$ (Player variance) 	& $1.675$ 	& 0.378 & $(1.085,2.519)$ \\
\hline
\end{tabular}
\end{table}

\newpage

\begin{table}[!hbt]
\caption{Comparison of four candidate models}
\centering
\label{tab:comparison}
\vspace{0.1in}
\begin{tabular}{llllll}
\hline
& & SLRM & KMM & NN & Our model \\
\hline
Headers & Brier Score  	& 0.091 	 & 0.101        &  0.089 & 0.061 \\
        & $-$log score  & 301.738    & 338.876      & 299.42  & 184.278 \\
        & error \% 	    & $11.03\%$  & $11.654\%$   & $11.03\%$  & $8.949\%$ \\
        & AUC           & 0.746      & 0.745        &  0.789     & 0.952 \\
\hline
Other shots & Brier Score  	& 0.094 	 & 0.098        &  0.097 & 0.067 \\
            & $-$log score  & 958.212    & 1043.567     & 976.625   & 633.371 \\
            & error \% 		& $11.983\%$ & $11.983\%$   & $12.216\%$  & $9.813\%$ \\
            & AUC           & 0.776      & 0.775        &  0.763     & 0.937 \\
\hline
\end{tabular}
\end{table}

\newpage

\begin{table}[!hbt]
\caption{Real-life statistics of best players as identified by our model}
\centering
\label{tab:sp-ps}
\vspace{0.1in}
\begin{tabular}{lccccc}
\hline
\multicolumn{6}{c}{Players with best SPS (ability to score with a shot)} \\
\hline
Player 	& SPS & PSS & Shots (on target) & Goals & Assists \\
\hline
Michael Barrios 	& 0.814	& 0.211	& 50 (26) & 9 & 2 \\
Ignacio Piatti 	    & 0.723	& 0.228	& 95 (42) & 17 & 6 \\
Giovani Dos Santos 	& 0.654 & 0.177	& 63 (31) & 14 & 12 \\
Giles Barnes 	    & 0.504	& 0.129	& 51 (19) & 6 & 3 \\
Kekuta Manneh 	    & 0.453 & 0.106	& 36 (20) & 5 & 2 \\
\hline
\multicolumn{6}{c}{Players with best PSS (positioning sense for shooting)} \\
\hline
Player & PSS & SPS & Shots (on target) & Goals & Assists \\
\hline
Ignacio Piatti 	    & 0.228	& 0.723	& 95 (42) & 17 & 6 \\
Fanendo Adi         & 0.219 & 0.124	& 88 (37) & 16 & 2 \\
Ola Kamara 		    & 0.216	& 0.178	& 80 (38) & 16 & 2 \\
Michael Barrios 	& 0.211	& 0.814	& 50 (26) & 9 & 2 \\
Giovani Dos Santos 	& 0.177 & 0.654	& 63 (31) & 14 & 12 \\
\hline
\multicolumn{6}{c}{Players with best SPH (ability to score with a header)} \\
\hline
Player 	& SPH & PSH & Shots (on target) & Goals & Assists \\
\hline
Drew Moor               & 0.937 & 0.208	& 21 (6)   & 3  & 0 \\
Bradley Wright-Phillips & 0.793 & 0.185	& 103 (56) & 24 & 5 \\
Cyle Larin              & 0.558 & 0.203	& 73 (33)  & 14 & 3 \\
Dominic Dwyer           & 0.537 & 0.152 & 119 (44) & 16 & 3 \\
Sebastian Hines         & 0.379 & 0.130 & 27 (9)   & 3 & 0 \\
\hline
\multicolumn{6}{c}{Players with best PSH (positioning sense for heading)} \\
\hline
Player & PSH & SPH & Shots (on target) & Goals & Assists \\
\hline
Drew Moor               & 0.208 & 0.937	 & 21 (6)   & 3  & 0 \\
Cyle Larin              & 0.203 & 0.558	 & 73 (33)  & 14 & 3 \\
Bradley Wright-Phillips & 0.185 & 0.793	 & 103 (56) & 24 & 5 \\
David Horst             & 0.154 & 0.150  & 21 (11)   & 3  & 0  \\
Dominic Dwyer           & 0.152 & 0.537  & 119 (44) & 16 & 3 \\
\hline
\end{tabular}
\end{table}

\newpage

\section{Figure captions}

\Cref{fig:heatmap}: Heat map of the full data: Total number of attempts were 3957, among which 482 were goals and 3475 were misses. Each pixel represents the proportion of goals scored from that position, color coded on a scale from 0\% (white) to 100\% (black).

\Cref{fig:shot-location}: (Left) For location X, the distance is the Euclidean distance between X and the center (O) of the goalline, while the angle (with appropriate sign) is $\angle XOY$, OY being the line that bisects the field horizontally. (Right) O is the ideal position of the goalkeeper (bisector of the angle the location of the shot makes with the goal), the line with small dots is the shot and OS is the keepreach, the shortest distance between the ideal location and the path of the shot.

\Cref{fig:rate-distance-angle}: Conversion rates corresponding to different distances (in yards) and angles (in degrees).

\Cref{fig:ripley}: Ripley's K-function $K(a)$ is plotted against $a$ for both headed and non-headed goals considering goals as spatial point processes. Under spatial homogeneity and independence, $K(a)=\pi a^2$. Expected values under this hypothesis and confidence intervals are constructed for reference.

\Cref{fig:score-cost}: Logarithm of the score function, corresponding to different cost values, for simple logistic regression model (SLRM), Klier-McMillen model (KMM), Neural network (NN) and for our proposed model.

\Cref{fig:histogram-sp-ps}: Histogram of the shooting prowess (left) and positioning senses (right) for 381 players, calculated based on fitted probability and actual outcomes.

\newpage

\section{Figures}





\newpage

\newpage

\newpage

\newpage

\newpage

\end{document}